\documentstyle[11pt,newpasp,twoside,epsf]{article}
\def\edcomment#1{\iffalse\marginpar{\raggedright\sl#1\/}\else\relax\fi}
\reversemarginpar

\begin{document}
\title{Direct Monte Carlo simulations of accretion discs}
\author{Takuya Matsuda, Hiromi Mizutani}
\affil{Department of Earth and Planetary Sciences, Kobe University,
Kobe, Japan}
\author{Henri M.J. Boffin}
\affil{Royal Observatory of Belgium, 3 av. Circulaire, 1180 Brussels}

\begin{abstract}
We apply the Direct Simulation Monte Carlo (DSMC) method,  developed originally to calculate rarefied gas dynamical problems, to study the gas flow in an accretion disc in a close binary system. The method involves viscosity and thermal conduction automatically, and can thus simulate turbulent viscosity ($\alpha$ viscosity). 
\end{abstract}

\section{Introduction}
The Direct Simulation Monte Carlo method was originally developed to treat rarefied gas flow (see Bird 1994; Matsuda, Mizutani \& Boffin 2001). In DSMC, microscopic molecular motions and their mutual collisions are calculated. Macroscopic quantities such as density, velocity and temperature of fluids are calculated by taking some mean of microscopic quantities. In DSMC, molecular collisions are handled in a stochastic manner, hence the Monte Carlo denomination of the method. The collision model in DSMC is such that the molecules are assumed to be hard billiard balls.
Since we solve the molecular motions, molecular viscosity and molecular thermal conduction are automatically included in the system. This 'physical' viscosity may mimic a turbulent viscosity ($\alpha$ viscosity) assumed in the standard theory of accretion disk. 

\section{Application of DSMC to a three-dimensional accretion disc}
We consider a binary system with mass ratio $q=0.5$. We work in the rotational frame. We restrict our computational region to a square, the center of which is the primary and just including the L1 point from which gas particles are injected constantly into the computational region. The region is divided into $100 \times 100 \times 20$ cells. The cell size is such that $dx=dy=2~dz$.
We start injection of gas particles at a constant rate per time step and follow the time evolution of the system. The number of particles in the computational region increases with time and reaches a steady state. After a steady state is reached, we further calculate to take a time average. Typically we computed up to t=60-100 in dimensionless unit, that is about 10-15 orbital periods.

In accretion discs, radiative cooling must occur, and some mechanism of cooling is essential in our calculation.  We here consider a simple model of cooling: we reduce the thermal speed of the molecules by a factor $f=0.97$ at each time step. 

\begin{figure}
\epsfxsize=8cm
%\centerline{\epsfbox{ae95_100bw.eps}}
\plottwo{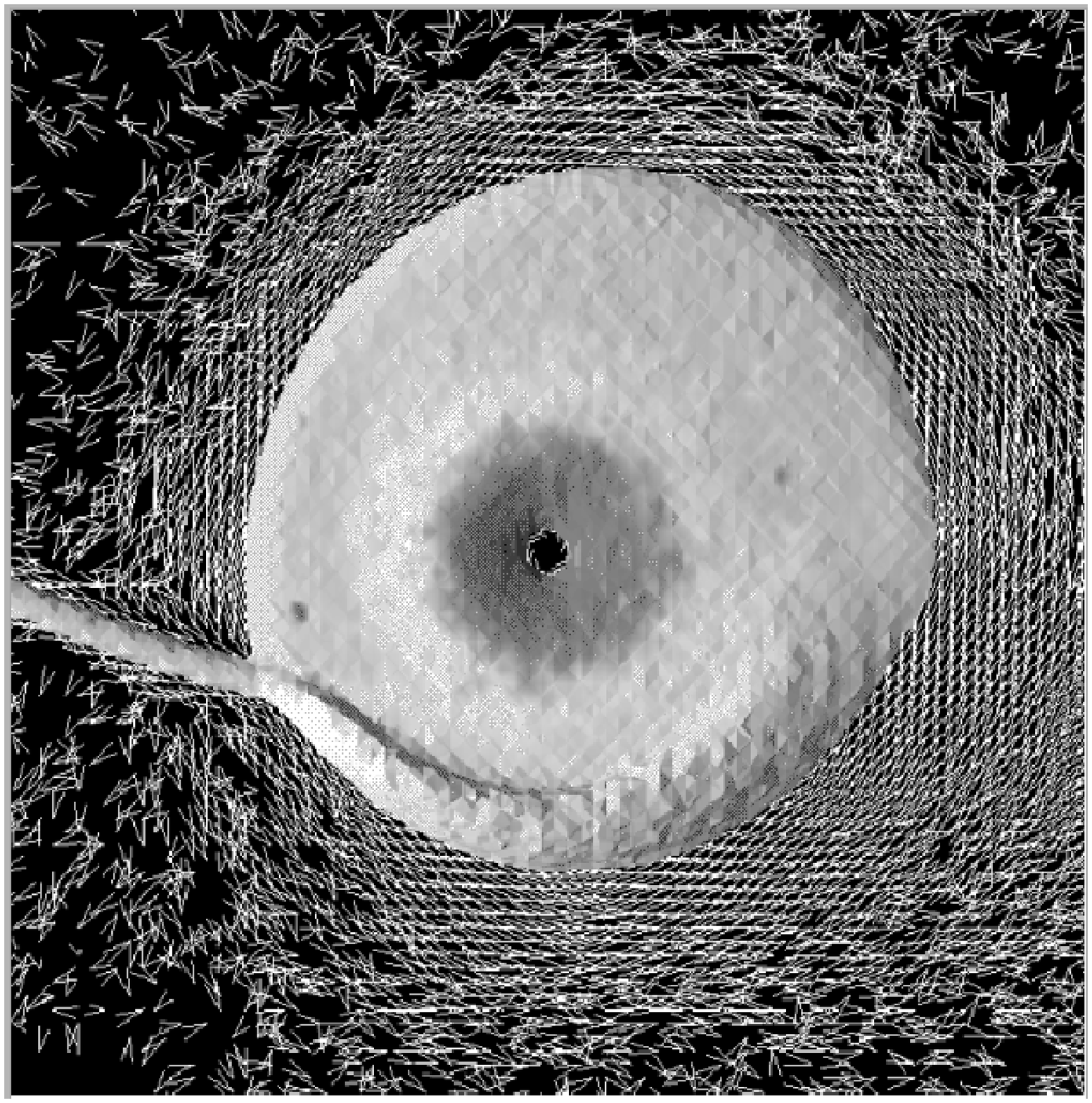}{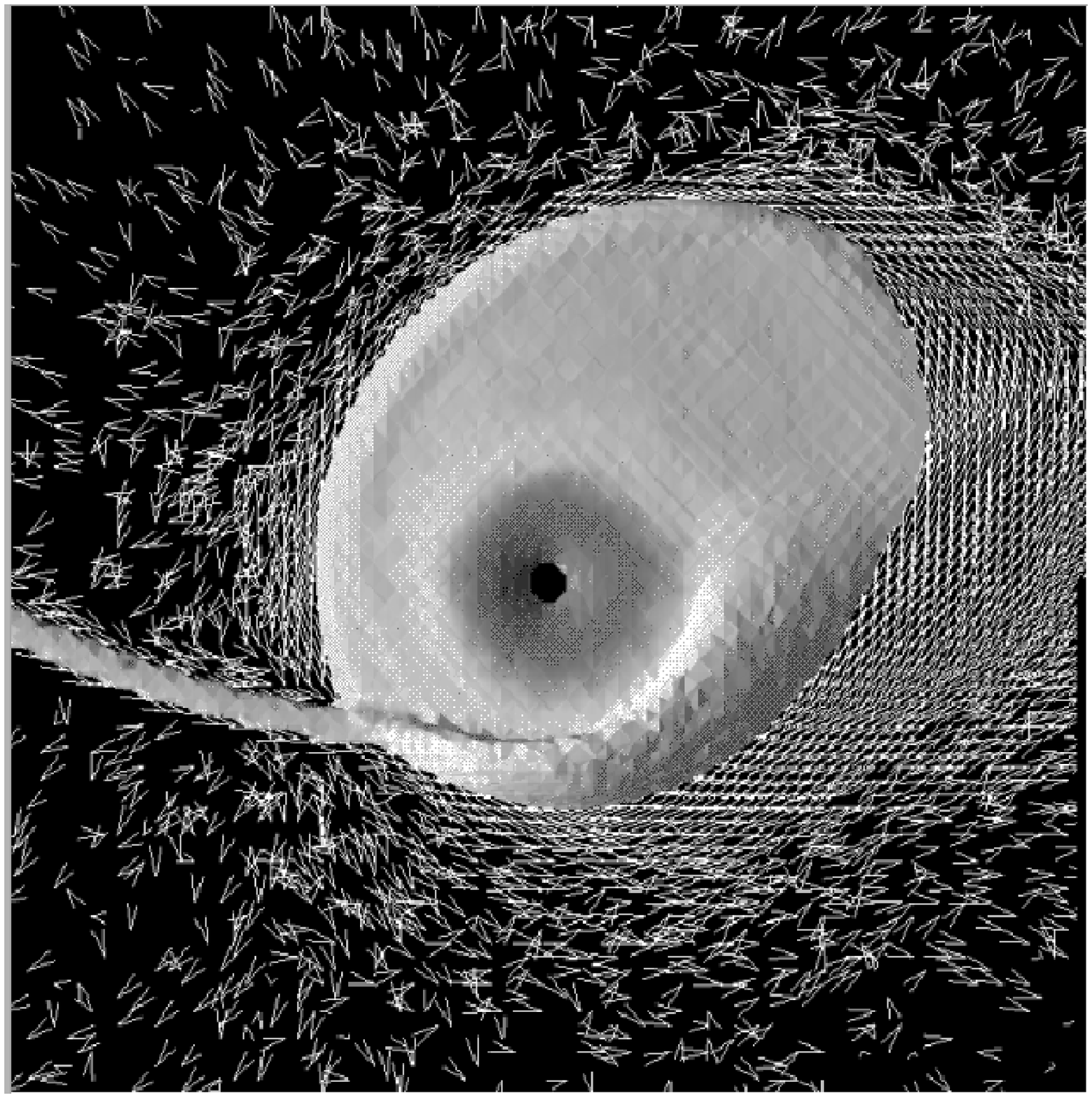}
\caption{Iso-density surface of an accretion disc with velocity vectors on the rotational plane; (left) 6 particles are injected every time step; (right) 30 particles are injected every time step}
\end{figure}

\section{Results}
Figure 1 displays the time averaged iso-density surface, on which temperature distribution is pasted,  and velocity vectors in the rotational plane of our simulation. Two cases of injected particles number, $n_{inj}=6$ and  $n_{inj}=30$, are compared. We can observe the gas stream from the L1 point, the accretion disc and spiral-shaped shocks. In the case of $n_{inj}=6$, we find that the accretion disc is very circular. The L1 stream hitting at the disc is deflected from its direction downward by the disc flow. Because of the collision between the L1 stream and the disc flow, the iso-density surface is elevated and the temperature on the elevated area is raised (a hot line or heated wall). In the case of $n_{inj}=30$, the L1 stream is so strong that the direction of the disc flow is deflected and the disc shape is elongated.  We may observe a pair of spiral shocks in this case.

\end{document}